\documentclass[a4paper, 10pt, conference]{ieeeconf}      

\IEEEoverridecommandlockouts    

\usepackage{amsmath,amssymb,amsfonts}
\usepackage{algorithmic}
\usepackage{graphicx}
\usepackage{textcomp}
\usepackage{xcolor}
\usepackage{multirow}
\newcommand\T{\rule{0pt}{2.6ex}}
\newcommand\B{\rule{0pt}{3.2ex}}

\usepackage{cite}
\usepackage{flushend}
\usepackage{tikz}

\def\BibTeX{{\rm B\kern-.05em{\sc i\kern-.025em b}\kern-.08em
    T\kern-.1667em\lower.7ex\hbox{E}\kern-.125emX}}

\title{ \Huge A Sensitivity-based Approach for Optimal Siting of Distributed Energy Resources}

\author{Mukesh Gautam*, \emph{Student Member, IEEE}, Narayan Bhusal*, \emph{Student Member, IEEE}, \\ Mohammed Benidris*, \emph{Member, IEEE}, Chanan Singh**, \emph{Fellow, IEEE}, and  Joydeep Mitra***, \emph{Fellow, IEEE}\\ *Department of Electrical \& Biomedical Engineering, University of Nevada, Reno,\\ (emails:  \{mukesh.gautam, bhusalnarayan62\}@nevada.unr.edu, and mbenidris@unr.edu)\\
**Department of Electrical \& Computer Engineering, Texas A\&M University, (email: singh@ece.tamu.edu)\\
***Department of Electrical \& Computer Engineering, Michigan State University, (email: mitraj@msu.edu)\vspace{-2.5ex}}

\begin{document}
\maketitle

\begin{abstract}
This paper presents a sensitivity-based approach for the placement of distributed energy resources (DERs) in power systems. The approach is based on the fact that most planning studies utilize some form of optimization, and solutions to these optimization problems provide insights into the sensitivity of many system variables to operating conditions and constraints. However, most of the existing sensitivity-based planning criteria do not capture ranges of effectiveness of these solutions (i.e., ranges of the effectiveness of Lagrange multipliers). The proposed method detects the ranges of effectiveness of Lagrange multipliers and uses them to determine optimal solution alternatives. Profiles for existing generation and loads, and transmission constraints are taken into consideration. The proposed method is used to determine the impacts of DERs at different locations, in presence of a stochastic element (load variability). This method consists of sequentially calculating Lagrange multipliers of the dual solution of the optimization problem for various load buses for all load scenarios. Optimal sizes and sites of resources are jointly determined in a sequential manner based on the validity of active constraints. The effectiveness of the proposed method is demonstrated through several case studies on various test systems including the IEEE reliability test system (IEEE RTS), the IEEE 14 and 30 bus systems. In comparison with conventional sensitivity-based approaches (i.e., without considering ranges of validity of Lagrange multipliers), the proposed approach provides more accurate results for active constraints.
\end{abstract}

\begin{keywords}
Lagrange multipliers, load variability, power system planning, sensitivity analysis.
\end{keywords}

\section{Introduction}
In recent years, integration of renewable energy sources as well as storage devices in the power grid has seen sustained growth. Conventional notions regarding limits on how much of these resources can be absorbed by the grid have been dispelled by numerous innovative approaches. This trend has motivated the development of innovative system planning methods that can foster and facilitate the integration of these resources. Determination of the optimal placement and sizes of these devices in terms of operation and ancillary services and participation in the electricity market is an important consideration in power system planning. Several methods have been introduced to solve such problems including analytical and population-based intelligent search methods. In this work, an analytical method that is based on the sensitivity analysis concept is proposed to jointly determine optimal locations and sizes of distributed energy resources (DERs) with respect to the desired objective function.

Several approaches have been presented in the literature to determine optimal sizes and sites of distributed generation and storage for various purposes. Authors of \cite{7403153} have proposed a method to determine optimal locations of virtual synchronous generators to provide an inertial response. In \cite{7091050}, an efficient analytical---optimal power flow (EA-OPF) based method---has been proposed for optimal location of distributed generation in distribution system. In \cite{4966230}, an iterative-analytical method has been proposed to determine optimal sizes and sites of distributed generation for radial distribution systems to reduce network losses. In \cite{6826880}, an analytical approach has been proposed to determine the sizes and sites of distributed generation on distribution systems for losses minimization. A two-stage sequential Monte Carlo simulation (MCS)-based stochastic strategy has been proposed in \cite{PES2020NB} to determine minimum sizes of movable energy resources (MERs) for service restoration and reliability enhancement. In this approach, spanning tree search algorithm for optimal network configuration, Dijkstra's shortest path algorithm for optimal routes to deploy MERs, and the traveling time of MERs are incorporated. A multi-objective optimization framework for sizing and siting of distributed generation based on genetic algorithm and an $\varepsilon$-constrained method has been proposed in \cite{1425569}. In \cite{5398822}, a reliability-based method has been used to determine the size of backup storage units. In \cite{7924418}, optimal locations of virtual synchronous generators have been proposed, which are determined based on an $H_2$ norm performance metric reflecting network coherency. An optimal size and location of battery energy storage system for load leveling has been proposed in \cite{486591}. 

A sensitivity analysis-based approach has been used in \cite{7747855} to determine optimal locations and sizes of DERs. The results have been validated using modified genetic algorithm. However, the method proposed in \cite{7747855} does not consider reactive power consumption nor does it consider dispatchable distributed energy resources; distributed energy resources are considered to produce only real power with ``must-take'' paradiem.  

In this paper, a sensitivity-based method is developed and applied to determine impacts of DERs placement at different locations considering load variability. The proposed approach is based on the fact that most planning studies utilize some form of optimization, and solutions to these optimization problems provide insights into the sensitivity of many system variables to operating conditions and constraints. However, most of existing sensitivity-based planning criteria do not capture ranges of effectiveness of these solutions (i.e., ranges of the effectiveness of Lagrange multipliers). The  proposed  method  detects  the  ranges  of  effectiveness of  Lagrange  multipliers  and  uses  them  to  determine  optimal solution alternatives. Profiles of existing generation and loads, and transmission constraints are taken into consideration. Since DERs significantly influence voltage profiles and reactive power requirements, these too are included in the optimization framework. Although several objective functions such as loss reduction, reliability maximization, power quality improvement, and cost minimization can be achieved using the proposed approach, this paper only  considers cost minimization. Other functions will be included in future work. The effectiveness of the proposed method is demonstrated through several case studies on various test systems including the IEEE reliability test system (IEEE RTS) and the IEEE 14 and 30 bus systems. The proposed approach provides more accurate results than conventional sensitivity-based approaches.

The rest of the paper is organized as follows. Section \ref{sensitivity}, \ref{approach}, \ref{algorithm}, and  \ref{network} provide, respectively, an overview of sensitivity analysis and Lagrange multiplier-based methods; proposed approach of considering ranges of validity of Lagrange multipliers;  a solution algorithm to proposed approach; and network modeling and power flow techniques. Section \ref{case} presents case studies and discussions. Section \ref{conclusion} provides concluding remarks.
 
\section{Sensitivity Analysis }\label{sensitivity} 
Sensitivity analysis is an effective tool to assess the effect of optimization problem constraint relaxations on the objective function. Lagrange multiplier-based sensitivity analysis has numerous applications in different areas. Lagrange multipliers have been first proposed by the economist Conte Petrovic in terms of shadow prices. In his work, the linear programming has been used to maximize the output of some products \cite{IEEE42009}. Lagrange multipliers have been defined from different perspectives in various literature. For instance, from a primal-dual perspective, it has been defined as the dual variables associated with the linear/nonlinear programming problem. From optimization point of view, it has been defined as the rate of change of an objective function for an infinitesimal change in the right-hand side of the optimization problem. From the geometric prospective, it has been explained as the sub-gradients of the objective function along the dimension of resource provisioning changes.

Several variations of sensitivity analysis have been used to determine the change in an objective function with respect to problem constraints. For instance, it has been used in \cite{4075996} to forecast the short-term transmission congestion. In \cite{9042569}, Lagrange multipliers-based sensitivity coefficients have been used for adaptive load shedding. Lagrange multipliers based sensitivity analysis has been used in \cite{387946} for power system reliability enhancement. Normalized Lagrange multipliers have been used in \cite{667350} for topology error identification. In \cite{260857}, a Lagrange relaxation technique has been used for scheduling of hydro-thermal power systems.  Authors of \cite{806185} have used an augmented Lagrange multiplier method to determine optimal locations of unified power flow controllers. In \cite{1626357}, Lagrange multipliers have been used to determine the marginal value of spinning reserve and the marginal value of interruptible load. Lagrange multipliers have been used for reliability optimization in \cite{5216020}. However, using Lagrange multipliers without determining their range of effectiveness could produce inaccurate results. This can be attributed in part to the fact that Lagrange multipliers change with the change in system conditions. For instance, buses that are ranked as the best candidates with respect to generation costs may not be valid for large energy sources or storage devices since Lagrange multipliers change with the change in system loading. In this paper, the Lagrange duality concept with considering ranges of validity of Lagrange multipliers has been used as a decision-making tool to determine the optimal sizes and location of DERs.

\section{The Proposed Method}\label{approach}
In this paper, the Lagrange multiplier-based sensitivity approach has been used as a decision-making tool to determine the optimal sizes and locations of DERs. The proposed method detects the ranges of the effectiveness of Lagrange multipliers and uses them to determine optimal solution alternatives. Profiles of existing generation and loads, and transmission constraints are taken into consideration. Apart from this, the stochasticity of the load has been included using non-sequential Monte Carlo simulation (MCS). For detecting the ranges of effectiveness of Lagrange multipliers, the proposed method checks the validity of the active constraints before finalizing each of the locations for the placement of DERs. 

An optimization problem in the standard form can be expressed as follows. 
\begin{equation} \label{eq1}
    \mbox{Minimize } f(x)\mbox{,}
\tag{1}\end{equation}
subject to 
\begin{gather*}
    g_{j}(x)\leq 0 ; j=1, ...., m\mbox{,}     \\
    \tag{2} \label{eqconst}
    h_{j}(x)=0 ; j=1, ...., p\mbox{.}
\end{gather*}
In (\ref{eq1}) and \eqref{eqconst}, $f(x)$ denotes objective function; $g_{j}(x)$ denotes inequality constraints; $h_{j} (x)$ denotes equality constraints; and $x \in \mathbb{R}^{n}$.

The basic idea in Lagrangian duality is to take the constraints in \eqref{eqconst} into account by augmenting the objective function of \eqref{eq1} with a weighted sum of the constraint functions\cite{cambridge73004}. The Lagrangian $L$: $\mathbb{R}^n\times \mathbb{R}^m \times \mathbb{R}^p \xrightarrow{} \mathbb{R}$ associated with the problem of \eqref{eq1} and \eqref{eqconst} can be defined as follows.
\begin{equation*}\label{eq:lag}
    L(x, \lambda, \nu)=f(x)+\sum_{j=1}^{m}\lambda_{j}g_{j}(x)+\sum_{j=1}^{p}\nu_{j}h_{j}(x) \tag{3}
\end{equation*}
where $\lambda_{j}$ refers to the Lagrangian multiplier associated with the $j$th 
inequality constraint: $g_{j}(x) \leq 0$; and $\nu_{j}$ is Lagrange multiplier associated with the $j$th equality constraint: $h_j(x)$.

The Lagrange dual function $\theta$: $\mathbb{R}^m\times\mathbb{R}^p\xrightarrow{}\mathbb{R}$ is denoted as the minimum value of the Lagrangian over $x$: for $\lambda \in \mathbb{R}^m$, $\nu \in \mathbb{R}^p $. Mathematically, it can be expressed as follows.
\begin{equation*}\label{eq:lagdual}
    \theta(\lambda, \nu)=\mbox{inf } \{L(x, \lambda, \nu)\}\tag{4}
\end{equation*}

\begin{figure}
   \vspace{-4ex}
    \hspace{-3.6ex}
    \includegraphics{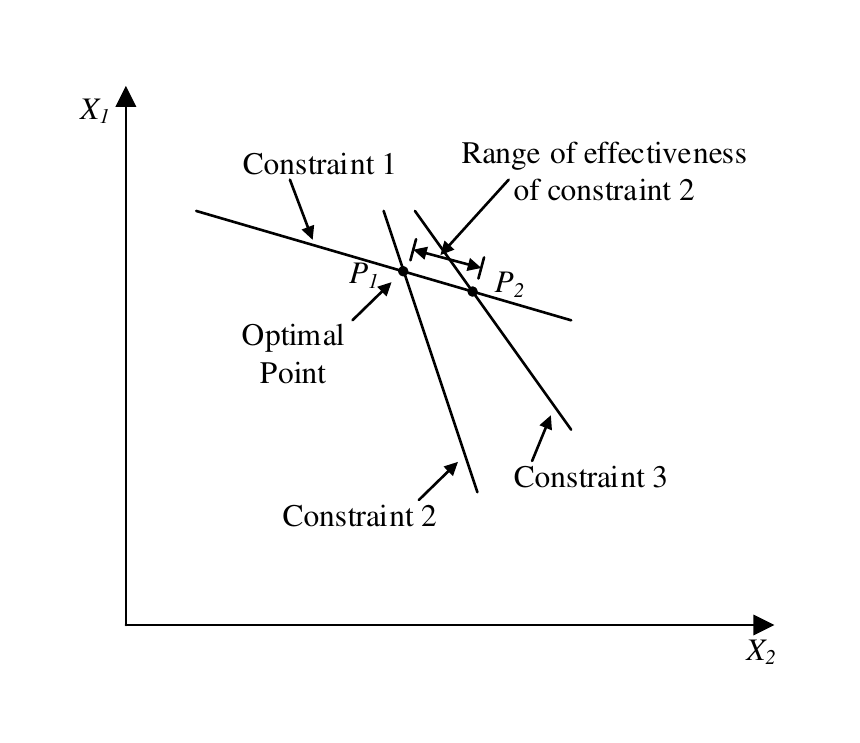}
    \vspace{-5ex}
    \caption{Change of Lagrange multipliers with loading and generation}
    \vspace{-1ex}
    \label{fig2}
\end{figure}

The Lagrange multipliers used in this paper are associated with the power balance equations which are also affected by the loading level of the system. To illustrate the change in values of Lagrange multipliers with the change in system conditions, consider a linear system with two variables ($X_1$ and $X_2$) and three constraints as shown in Fig.\ref{fig2}.

It can be seen from Fig.\ref{fig2} that initially when the system is operating at the optimal point $P_1$, constraint $3$ is inactive, thus the Lagrange multiplier associated with constraint $3$ is zero. During this condition, the optimal operating point is constrained by constraint $1$ and constraint $2$ only. The constraint $2$ can still be relaxed up to point $P_2$ without making it inactive. If constraint $2$ is further relaxed beyond point $P_2$, it becomes inactive and constraint $3$ becomes active. Here, $P_1$ to $P_2$ is the range of effectiveness of constraint $2$.

In the proposed method, the weighted average values and range of validity of Lagrange multipliers obtained after solving the optimization problem for several iterations under variable loading conditions are used to determine the locations that are more sensitive to load variation. Due to this reason, the DER with the highest capacity is placed in the location that has the highest weighted average Lagrange multiplier.

\section{Solution Algorithm}\label{algorithm}
This section describes the solution algorithm to the proposed Lagrange multiplier-based sensitivity approach for placement of DERs. In the proposed method, MCS is used for determining the average values of Lagrange multipliers. In MCS, the states are sampled from the state space proportional to their probabilities \cite{7478112}.
Moreover, it is easy to implement and requires less computation time. 
A stopping criterion is required to stop the simulation after the convergence of Lagrange multipliers \cite{7302681}. In this paper, the stopping criterion applied on Lagrange multipliers is calculated as follows.
\begin{gather*}
    \sigma = \text{max}\Bigg(\frac{\sqrt{\text{Var}(\lambda_k)}}{E[\lambda_k]}\Bigg) ; k=1, ...., N_b
    \tag{5}
\end{gather*}
where $E$[.] is the expectation operator; Var(.) is the variance operator; $\lambda_k$ is Lagrange multiplier of bus $k$; and $N_b$ is the number of buses in the system.

The solution algorithm to system planning using the proposed method can be explained as follows.

\begin{enumerate}
    \item Read system data (such as bus data, branch data and generator data) along with their hourly load profile.
    \item Using the hourly load profile for a year, cluster load levels into $50$ clusters (any number of clusters can be used based on the required accuracy) along with their cumulative probability distribution.
    \item Start MCS and generate a random number. Select the load level corresponding to the generated random number.
    
    \item Using the sampled value of the load level, solve the optimization problem. For the dual solution of the optimization problem, compute Lagrange multipliers for each node of the test system associated with power flow equations. Rank Lagrange multipliers in a descending order, which are used to select the nodes for DERs to be added.
    
    \item According to the number of DERs to be added, their sizes and ranked Lagrange multipliers, determine the locations and sizes for DERs placement. For example, if three DERs of different sizes are to be placed, the highest three Lagrange multipliers are used for determining the sizes and locations. This implies that the highest capacity DER is placed in the location with the highest value of Lagrange multiplier.
    
    \item The optimization problem is again solved after DERs placement and new values of Lagrange multipliers are computed. Check the inactivity of power flow constraints at each location. Penalize the bus(es) in which power flow constraints are inactive, since those locations are no longer responsible for optimality of the solution. Store these new values of the Lagrange multipliers. 
    
    \item Check whether the convergence criterion is met. If yes, go to the next step. Otherwise, generate a new random number, select the load level corresponding to the generated random number and go back to step $4$. 
    
    \item Compute the weighted average value of the Lagrange multipliers from the stored values of the Lagrange multipliers during each iteration of MCS. Based on these values of Lagrange multipliers, determine the locations of DERs to be integrated.
\end{enumerate}

\begin{figure}
    \vspace{-2ex}
    \includegraphics[scale=0.85]{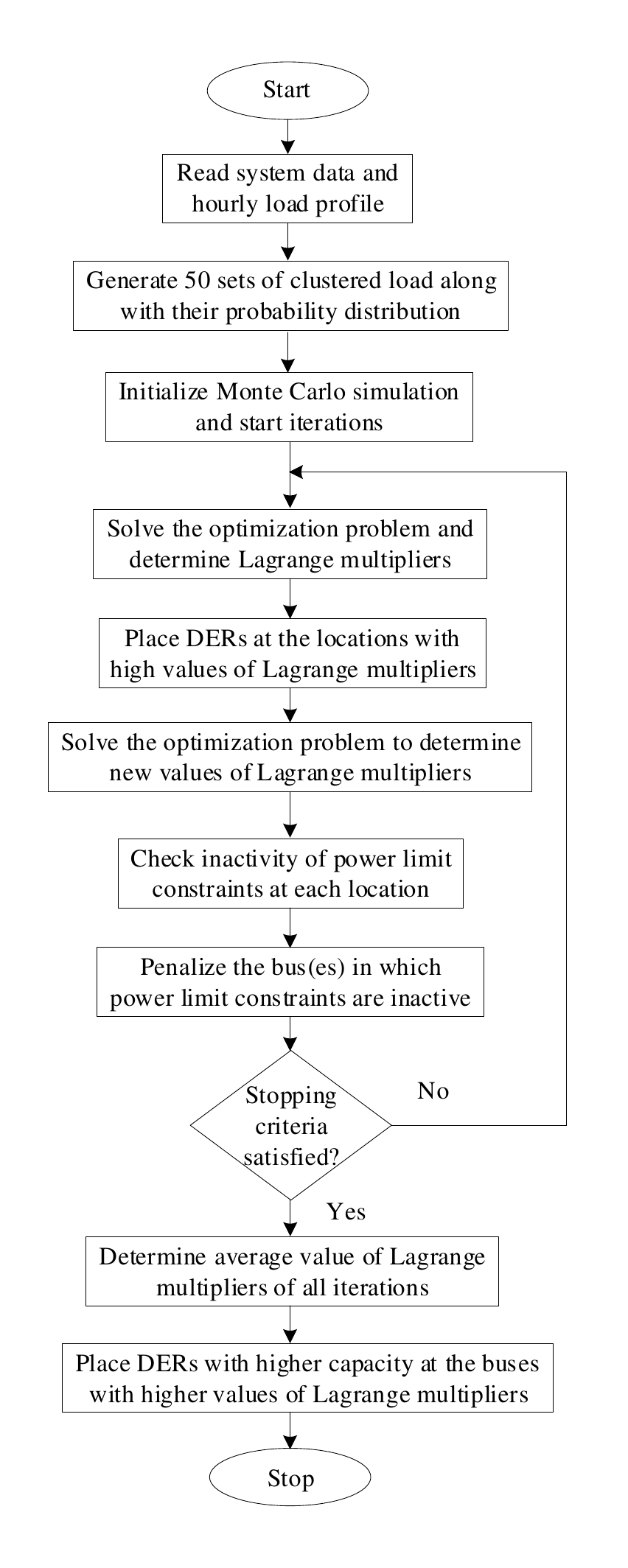}
    \vspace{-3ex}
    \caption{Flow chart of proposed solution algorithm}
    \label{fig1}
    \vspace{-1ex}
\end{figure}

The flow chart of the proposed solution algorithm is shown in Fig. \ref{fig1}.   

\section{Network Modeling and Power Flow}
In this paper, the AC optimal power flow model is used to solve the optimization problem for the determination of Lagrange multipliers. For solving the optimal power flow, generation cost minimization is used as the objective function. The objective function is subjected to equality and inequality constraints of power system operation limits. The equality constraints include power balance equations at each bus, and the inequality constraints include capacity limits of each generating unit, voltage limits at each bus and reactive power capability limits. The network model can be formulated as follows.
\label{network}
\begin{equation*} \label{eq2}
C=\mbox{min}\sum_{j=1}^{N_g}C_{j}(P_{Gj})\mbox{,}
\tag{6}\end{equation*}
subject to 
\begin{gather*} P(V, \ \delta)-P_{D}=0\mbox{,}\\ Q(V, \ \delta)-Q_{D}=0\mbox{,}\\ P_{G}^{min}\leq P(V, \ \delta)\leq P_{G}^{max}\mbox{,}\\ Q_{G}^{min}\leq Q(V, \ \delta)\leq Q_{G}^{max} \tag{7}\mbox{,}\\ V^{min}\leq V\leq V^{max}\mbox{,}\\ S(V, \ \delta)\leq S^{\max}\mbox{,}\\ \delta\ \text{unrestricted}\mbox{,} \end{gather*}
where
\begin{gather*}
    \begin{aligned}
    &C :~ \text{the generation cost function;}  \\
    &N_g:~ \text{total number of generating units;} \\
   &C_j:~\text{the generation cost function of generating unit j;} \\
   &P_D:~\text{the active power demand vector} (N_d \times 1);\\
   &Q_D:~\text{the reactive power demand vector} (N_d \times 1);\\
    &V:~ \text{the bus voltage magnitude vector}(N_b \times 1);\\
   &\delta:~ \text{the bus voltage angles vector}(N_b \times 1);\\
   &V^{min} \& V^{max}:~ \text{the bus voltage limits vector}(N_b \times 1);\\
   &V:~ \text{the bus voltage magnitude vector}(N_b \times 1);\\
   &P(V, \delta):~\text{the active power injection vector}(N_b \times 1);\\
   &Q(V, \delta):~\text{the reactive power injection vector}(N_b \times 1);\\
   &S(V, \delta):~\text{the line flow vector}(N_t \times 1);\\
   &S^{max}:~\text{the line flow limits vector}(N_t \times 1);\\
   &N_b:~ \text{total number of buses};\\
   &N_d:~ \text{total number of load buses};\\
   &N_t:~ \text{total number of transmission lines};\\
   &P_{G}^{min}:~ \text{minimum active power generation limit}; \\
   &P_{G}^{max}:~ \text{maximum active power generation limit}; \\
   &Q_{G}^{min}:~ \text{minimum reactive power generation limit};\\
   &Q_{G}^{max}:~ \text{maximum reactive power generation limit}.
   \end{aligned}
\end{gather*}

\vspace{1ex}

\section{Case Studies and Discussions}\label{case}
The proposed method and solution algorithm is demonstrated on several test systems including the IEEE 14 bus system, IEEE 30 bus system, and IEEE RTS. These systems have been extensively used for optimal placement and sizing of DERs for various research objectives \cite{7747855}. The IEEE 14 bus system consists of 14 buses, 5 generators, and 11 loads with total generation capacity of $772.4$ MW and total peak load of $259$ MW. The IEEE 30 bus system consists of 30 buses, 6 generators, and 20 loads with total generation capacity of $335$ MW and total peak load of $189.2$ MW. The detailed data of IEEE 14 and IEEE 30 bus system are given in \cite{ieee14bus} and \cite{4075418}, respectively. The IEEE RTS consists of 24 buses, 33 transmission line, 5 transformers, and 32 generating units on 10 buses with total generation capacity of $3405$ MW and peak load of $2850$ MW. The detailed data (e.g. size and type of generators, failure rate of the various system components, and load profile) of IEEE RTS are given in \cite{4113721}. 

Before performing MCS, the system load is modeled by a clustering technique where 50 clusters are generated along with their cumulative probability distribution. By drawing a uniformly distributed random number, the load level can be determined according to the position of the random number. This load level is used to determine actual load at each bus for all test systems.

The optimal AC power flow is solved using MATPOWER (open source, version 7.0) \cite{5491276}. Four case studies are performed for different operation paradigms of DERs (dispatchable and non-dispatchable real and reactive power) on the aforementioned test systems. We performed these case studies to test the proposed approach and ranges of validity of Lagrange multipliers with change in real and reactive power control.

\subsection{Case \ref{tab1}: Without considering the range of validity of active constraints}
In this case, the DERs with non-dispatchable active and reactive power are considered, and the range of the validity of active constraints is not considered. This case study is performed to compare the proposed approach which considers ranges of validity of Lagrange multipliers with conventional sensitivity-based methods. Table \ref{tab1} shows the results for this case. The results for the IEEE $14$ bus system show that bus $3$ is most sensitive to the change in load. So, the DER with the highest capacity is placed at bus $3$. At buses $10$ and $9$, small DERs are placed in descending order of their average Lagrange multipliers.

\begin{table}[h!]
\caption{Optimal Locations and Sizes of DERs without considering the range of validity of active constraints\vspace{-1.5ex}}
\begin{center}
\begin{tabular}{|c|c|c|c|c|}
\hline
\multicolumn{2}{|c|}{Systems} & \multicolumn{3}{c|}{Optimal Locations and Sizes} \B\\\hline
\multicolumn{1}{|c|}{\multirow{2}{*}{IEEE 14 bus}} & \multicolumn{1}{c|}{Bus} & \multicolumn{1}{c|}{3} & \multicolumn{1}{c|}{10} & \multicolumn{1}{c|}{9} \T\\ \cline{2-5} 
\multicolumn{1}{|c|}{} & \multicolumn{1}{c|}{Size*} & \multicolumn{1}{c|}{$30;10$} & \multicolumn{1}{c|}{$20;6.66$} & \multicolumn{1}{c|}{$10;3.33$} \T\\ \hline
\multicolumn{1}{|c|}{\multirow{2}{*}{IEEE 30 bus}} & \multicolumn{1}{c|}{Bus} & \multicolumn{1}{c|}{8} & \multicolumn{1}{c|}{21} & \multicolumn{1}{c|}{17} \T\\ \cline{2-5} 
\multicolumn{1}{|c|}{} & \multicolumn{1}{c|}{Size*} & \multicolumn{1}{c|}{$30;10$} & \multicolumn{1}{c|}{$20;6.66$} & \multicolumn{1}{c|}{$10;3.33$} \T\\ \hline
\multicolumn{1}{|c|}{\multirow{2}{*}{IEEE RTS}} & \multicolumn{1}{c|}{Bus} & \multicolumn{1}{c|}{4} & \multicolumn{1}{c|}{5} & \multicolumn{1}{c|}{2} \T\\ \cline{2-5} 
\multicolumn{1}{|c|}{} & \multicolumn{1}{c|}{Size*}& \multicolumn{1}{c|}{$60;20$} & \multicolumn{1}{c|}{$50;16.5$} & \multicolumn{1}{c|}{$40;13.2$} \T\\ \hline
\multicolumn{5}{l}{*: $P$ (MW)$;$ Q (MVar)}\T\\
\end{tabular}
\label{tab1}
\end{center}
\end{table}

\subsection{Case \ref{tab2}: The proposed method with non-dispatchable active and reactive power}
In this case, the DERs with non-dispatchable active and reactive power are considered and the proposed method is applied for all of the aforementioned test systems. Table \ref{tab2} shows the results of the proposed method for DERs with non-dispatchable active and reactive power. The results show that when the range of validity of the active constraints is considered, the optimal locations of DERs may change for the locations which have short range of validity of their active constraints. The optimal locations of DERs for IEEE $30$ bus system have not changed but the optimal locations for IEEE $14$ bus system and IEEE RTS have changed. This implies that the optimal locations of DERs for IEEE $30$ bus system have large range of validity of their active constraints.  
\begin{table}[h!]
\caption{Optimal Locations and Sizes of DERs with non-dispatchable active and reactive power\vspace{-1.5ex}}
\begin{center}
\begin{tabular}{|c|c|c|c|c|}
\hline
\multicolumn{2}{|c|}{Systems} & \multicolumn{3}{c|}{Optimal Locations and Sizes} \B\\\hline
\multicolumn{1}{|c|}{\multirow{2}{*}{IEEE 14 bus}} & \multicolumn{1}{c|}{Bus} & \multicolumn{1}{c|}{10} & \multicolumn{1}{c|}{9} & \multicolumn{1}{c|}{7} \T\\ \cline{2-5} 
\multicolumn{1}{|c|}{} & \multicolumn{1}{c|}{Size*} & \multicolumn{1}{c|}{$30;10$} & \multicolumn{1}{c|}{$20;6.66$} & \multicolumn{1}{c|}{$10;3.33$} \T\\ \hline
\multicolumn{1}{|c|}{\multirow{2}{*}{IEEE 30 bus}} & \multicolumn{1}{c|}{Bus} & \multicolumn{1}{c|}{8} & \multicolumn{1}{c|}{21} & \multicolumn{1}{c|}{17} \T\\ \cline{2-5} 
\multicolumn{1}{|c|}{} & \multicolumn{1}{c|}{Size*} & \multicolumn{1}{c|}{$30;10$} & \multicolumn{1}{c|}{$20;6.66$} & \multicolumn{1}{c|}{$10;3.33$} \T\\ \hline
\multicolumn{1}{|c|}{\multirow{2}{*}{IEEE RTS}} & \multicolumn{1}{c|}{Bus} & \multicolumn{1}{c|}{14} & \multicolumn{1}{c|}{7} & \multicolumn{1}{c|}{8} \T\\ \cline{2-5} 
\multicolumn{1}{|c|}{} & \multicolumn{1}{c|}{Size*}& \multicolumn{1}{c|}{$60;20$} & \multicolumn{1}{c|}{$50;16.5$} & \multicolumn{1}{c|}{$40;13.2$} \T\\ \hline
\multicolumn{5}{l}{*: $P$ (MW)$;$ Q (MVar)}\T\\
\end{tabular}
\label{tab2}
\end{center}
\end{table}
\vspace{-3ex}
\subsection{Case \ref{tab3}: The proposed method with dispatchable reactive power and non-dispatchable active power}
In this case, optimal locations and sizes are determined for DERs with dispatchable reactive power but non-dispatchable active power. Table \ref{tab3} shows the results of the proposed method for DERs with dispatchable reactive power but non-disptchable active power. The results show that the different optimal locations may be obtained when DERs with dispatchable reactive power is considered. Compared to case \ref{tab2}, the optimal locations of DERs for IEEE $14$ bus system have not changed but the optimal locations for IEEE $30$ bus system and IEEE RTS have changed. Again, this is because of the large range of validity of the active constraints of IEEE $30$ bus system.   

\begin{table}[h!]
\caption{Optimal Locations and Sizes of DERs with dispatchable reactive power and non-dispatchable active power\vspace{-1.5ex}}
\begin{center}
\begin{tabular}{|c|c|c|c|c|}
\hline
\multicolumn{2}{|c|}{Systems} & \multicolumn{3}{c|}{Optimal Locations and Sizes} \B\\\hline
\multicolumn{1}{|c|}{\multirow{2}{*}{IEEE 14 bus}} & \multicolumn{1}{c|}{Bus} & \multicolumn{1}{c|}{10} & \multicolumn{1}{c|}{9} & \multicolumn{1}{c|}{7} \T\\ \cline{2-5} 
\multicolumn{1}{|c|}{} & \multicolumn{1}{c|}{Size*} & \multicolumn{1}{c|}{$30;10$} & \multicolumn{1}{c|}{$20;6.66$} & \multicolumn{1}{c|}{$10;3.33$} \T\\ \hline
\multicolumn{1}{|c|}{\multirow{2}{*}{IEEE 30 bus}} & \multicolumn{1}{c|}{Bus} & \multicolumn{1}{c|}{19} & \multicolumn{1}{c|}{20} & \multicolumn{1}{c|}{18} \T\\ \cline{2-5} 
\multicolumn{1}{|c|}{} & \multicolumn{1}{c|}{Size*} & \multicolumn{1}{c|}{$30;10$} & \multicolumn{1}{c|}{$20;6.66$} & \multicolumn{1}{c|}{$10;3.33$} \T\\ \hline
\multicolumn{1}{|c|}{\multirow{2}{*}{IEEE RTS}} & \multicolumn{1}{c|}{Bus} & \multicolumn{1}{c|}{14} & \multicolumn{1}{c|}{7} & \multicolumn{1}{c|}{18} \T\\ \cline{2-5} 
\multicolumn{1}{|c|}{} & \multicolumn{1}{c|}{Size*}& \multicolumn{1}{c|}{$60;20$} & \multicolumn{1}{c|}{$50;16.5$} & \multicolumn{1}{c|}{$40;13.2$} \T\\ \hline
\multicolumn{5}{l}{*: $P$ (MW)$;$ Q (MVar)}\T\\
\end{tabular}
\label{tab3}
\end{center}
\end{table}

\subsection{Case \ref{tab4}: The proposed method with dispatchable active and reactive power}
In this case, optimal locations and sizes are determined for DERs with dispatchable active and reactive power. Table \ref{tab4} shows the results of the proposed method for DERs with dispatchable active and reactive power. The results of this case show that the optimal locations differ when DERs with dispatchable active and reactive power are considered. This is mainly because of the decrease in the range of validity of active constraints when both active and reactive power are controllable.

\begin{table}[h!]
\caption{Optimal Locations and Sizes of DERs with dispatchable active and reactive power\vspace{-1.5ex}}
\begin{center}
\begin{tabular}{|c|c|c|c|c|}
\hline
\multicolumn{2}{|c|}{Systems} & \multicolumn{3}{c|}{Optimal Locations and Sizes} \B\\\hline
\multicolumn{1}{|c|}{\multirow{2}{*}{IEEE 14 bus}} & \multicolumn{1}{c|}{Bus} & \multicolumn{1}{c|}{14} & \multicolumn{1}{c|}{10} & \multicolumn{1}{c|}{9} \T\\ \cline{2-5} 
\multicolumn{1}{|c|}{} & \multicolumn{1}{c|}{Size*} & \multicolumn{1}{c|}{$30;10$} & \multicolumn{1}{c|}{$20;6.66$} & \multicolumn{1}{c|}{$10;3.33$} \T\\ \hline
\multicolumn{1}{|c|}{\multirow{2}{*}{IEEE 30 bus}} & \multicolumn{1}{c|}{Bus} & \multicolumn{1}{c|}{30} & \multicolumn{1}{c|}{29} & \multicolumn{1}{c|}{19} \T\\ \cline{2-5} 
\multicolumn{1}{|c|}{} & \multicolumn{1}{c|}{Size*} & \multicolumn{1}{c|}{$30;10$} & \multicolumn{1}{c|}{$20;6.66$} & \multicolumn{1}{c|}{$10;3.33$} \T\\ \hline
\multicolumn{1}{|c|}{\multirow{2}{*}{IEEE RTS}} & \multicolumn{1}{c|}{Bus} & \multicolumn{1}{c|}{8} & \multicolumn{1}{c|}{4} & \multicolumn{1}{c|}{5} \T\\ \cline{2-5} 
\multicolumn{1}{|c|}{} & \multicolumn{1}{c|}{Size*}& \multicolumn{1}{c|}{$60;20$} & \multicolumn{1}{c|}{$50;16.5$} & \multicolumn{1}{c|}{$40;13.2$} \T\\ \hline
\multicolumn{5}{l}{*: $P$ (MW)$;$ Q (MVar)}\T\\
\end{tabular}
\label{tab4}
\end{center}
\end{table}

\section{Conclusion}\label{conclusion}
This paper has introduced a sensitivity-based method to determine optimal locations and sizes of DERs. By analyzing the impacts of DERs at different locations, sizes and locations of DERs is determined. In developing the proposed approach, the following variables have been taken into consideration: impacts of different DERs; profiles for existing generation and loads; and transmission constraints. Moreover, the load variablity was also considered for analyzing the impacts of load variations on Lagrange multipliers. The proposed method was demonstrated on several test systems including the IEEE RTS, the IEEE 14 and 30 bus systems. As the objective of the proposed approach is to minimize the generation cost, the optimal locations and sizes will result is the saving of the generation cost. In our future work we will implement this approach for other objectives such as reliability maximization, loss minimization, contribution in the electricity markets, and providing other ancillary services.

\bibliographystyle{IEEEtran}
\bibliography{References.bib}
\end{document}